\documentclass[twocolumn,showpacs,preprintnumbers,amsmath,aps]{revtex4}
\usepackage{graphicx}
\usepackage{dcolumn}
\usepackage{bm}
\usepackage{color}
\usepackage{multirow}

\begin{document}
\def\eq#1{(\ref{#1})}
\def\fig#1{Fig.\hspace{1mm}\ref{#1}}
\def\tab#1{\hspace{1mm}\ref{#1}}
\title{Characteristics of superconducting state in vanadium:\\ the Eliashberg equations and semi-analytical formulas}
\author{E. A. Drzazga$^{\left(1\right)}$}
\email{edrzazga@wip.pcz.pl}
\author{I. A. Domagalska$^{\left(1\right)}$}
\author{M. W. Jarosik$^{\left(1\right)}$}
\author{R. Szcz{\c{e}}{\'s}niak$^{\left(1\right)}$}
\author{J. K. Kalaga$^{\left(2\right)}$}
\affiliation{$^1$ Institute of Physics, Cz{\c{e}}stochowa University of Technology, Ave. Armii Krajowej 19, 42-200 Cz{\c{e}}stochowa, Poland}
\affiliation{$^2$ Quantum Optics and Engineering Division, Faculty of Physics and Astronomy, University of Zielona G{\'o}ra, Prof. Z. Szafrana 4a, 65-516 Zielona G{\'o}ra, Poland}
\date{\today}
\begin{abstract}
The superconducting state in vanadium characterizes with the critical temperature ($T_{c}$) equal to $5.3$~K. 
The Coulomb pseudopotential, calculated with the help of the Eliashberg equations, possesses anomalously high value
$\mu^{\star}\left(3\Omega_{\rm max}\right)=0.259$ or $\mu^{\star}\left(10\Omega_{\rm max}\right)=0.368$ 
($\Omega_{\rm max}$ denotes the maximum phonon frequency). Despite the relatively large electron-phonon coupling constant ($\lambda=0.91$), the quantities such as: the order parameter ($\Delta$), the specific heat ($C$), and the thermodynamic critical field ($H_{c}$) determine the values of the dimensionless ratios not deviating much from the predictions of the BCS theory: 
$R_{\Delta}=2\Delta\left(0\right)\slash k_{B}T_{c}=3.68$, 
$R_{C}=\Delta C\left(T_{c}\right)\slash C^{N}\left(T_{c}\right)=1.69$, and 
$R_{H}=T_{c}C^{N}\left(T_{c}\right)\slash H^{2}_{c}\left(0\right)=0.171$. This result is associated with the reduction 
of the strong-coupling and the retardation effects by the high value of the Coulomb pseudopotential. 
It has been shown that the results of the Eliashberg formalism can be relatively precisely reproduced with the help 
of the semi-analytical formulas, if the value of $\mu^{\star}$ is determined on the basis of the $T_{c}$-Allen-Dynes expression 
($\mu^{\star}_{AD}=0.198$). The attention should be paid to the fact that in the numerical and in the semi-analytical approach the comparable values of the thermodynamic parameters for the same $\mu^{\star}$ have been obtained only in the vicinity of the point $\mu^{\star}=0.1$.     
\end{abstract}
\maketitle
\vspace*{1cm}

The classical Eliashberg theory is used to quantitative description of the superconducting state, which is mediated by the electron-phonon interaction \cite{Eliashberg1960A}. The strong-coupling and the retardation effects omitted in BCS model \cite{Bardeen1957A, Bardeen1957B} are included in the Eliashberg theory by assumption that the electron band energy is directly renormalized by the frequency-dependent order parameter ($\Delta_{n}=\Delta\left(i\omega_{n}\right)$), the wave function renormalization factor ($Z_{n}=Z\left(i\omega_{n}\right)$), and the energy shift function ($\chi_{n}=\chi\left(i\omega_{n}\right)$). The symbol $\omega_{n}=\left(\pi\slash\beta\right)\left(2n-1\right)$ denotes the Matsubara frequency, $\beta=1\slash k_{B}T$, where $k_{B}$ is the Boltzmann constant. In most cases, the materials, in which we observe the phonon induced superconducting state, characterize with the very wide electron band ($W\gg\Omega_{\rm max}$, where $W$ denotes the half-width of the band, and $\Omega_{\rm max}$ is the Debye frequency). For this reason, the analysis of the superconducting state can be conducted with the help of the Eliashberg equations for the half-filled electron band \cite{Carbotte1990A, Carbotte2003A}:
\begin{equation}
\label{r1}
\phi_{n}=\frac{\pi}{\beta}\sum_{m=-M}^{M}
\frac{\lambda\left(i\omega_{n}-i\omega_{m}\right)-\mu^{\star}\theta\left(\omega_{c}-|\omega_{m}|\right)}
{\sqrt{\omega_m^2Z^{2}_{m}+\phi^{2}_{m}}}\phi_{m},
\end{equation}
\begin{equation}
\label{r2}
Z_{n}=1+\frac{1}{\omega_{n}}\frac{\pi}{\beta}\sum_{m=-M}^{M}
\frac{\lambda\left(i\omega_{n}-i\omega_{m}\right)}{\sqrt{\omega_m^2Z^{2}_{m}+\phi^{2}_{m}}}\omega_{m}Z_{m},
\end{equation}
whereas the order parameter is given by the formula $\Delta_{n}={\phi_{n}}\slash{Z_{n}}$. The electron-phonon pairing kernel has the form of $\lambda\left(i\omega_{n}-i\omega_{m}\right)=2\int_0^{+\infty}d\omega\frac{\alpha^{2}F\left(\omega\right)\omega}
{\left(\omega_{n}-\omega_{m}\right)^2+\omega ^2}$. The spectral function $\alpha^{2}F\left(\omega\right)$ is usually calculated numerically by using the {\it ab initio} approach, where the electron band energy, the phonon dispersion relation, and the electron-phonon matrix elements are taken into consideration in the possible accurate way \cite{Baroni1986A, Giannozzi2009A}. Experimentally, $\alpha^{2}F\left(\omega\right)$ is obtained from the McMillan-Rowell tunnelling inversion procedure \cite{McMillan1965A}, which additionally allows to confirm the results of {\it ab initio} calculations. The depairing processes in the Eliashberg formalism are modelled parametrically ($\mu^{\star}$). The symbol $\theta$ represents the Heavisides unit function and $\omega_{c}$ is the cut-off frequency. For the low value of $\mu^{\star}$ ($\sim 0.1$), the considered parameter is called as the Coulomb pseudopotential and represents the Coulomb static screened repulsion corrected by the retardation effects \cite{Morel1962A, Bauer2012A}. In the case when 
$\mu^{\star}$ assumes significantly higher values, the considered quantity models in the most easy way the additional depairing processes, inter alia: the existence of the competing phases (CDW or SDW), the strong spin fluctuations, or the non-adiabatic effects. 

The Eliashberg set cannot be solved analytically. For this reason, the semi-analytical expressions exist in the literature, which enable to calculate the critical temperature ($T_{c}$), the order parameter at zero Kelvin ($\Delta\left(0\right)$), the superconducting and the normal specific heat 
($C^{S}$ and $C^{N}$), or the thermodynamic critical field ($H_{c}$). Below we openly present the formula for the critical temperature obtained by Allen and Dynes (known as the AD formula) \cite{Allen1975A}:
\begin{equation}
\label{r3}
k_{B}T_{c}=f_{1}f_{2}\frac{\omega_{\ln}}{1.2}\exp\left[\frac{-1.04\left(1+\lambda\right)}
{\lambda-\mu^{\star}\left(1+0.62\lambda\right)}\right],
\end{equation}
where the phonon logarithmic frequency possesses the form:
$\omega_{\ln}=\exp\left[\frac{2}{\lambda}\int^{+\infty}_{0}d\omega\frac{\alpha^{2}F\left(\omega\right)}{\omega}\ln\left(\omega\right)\right]$. 
The electron-fonon coupling constant can be written as:
$\lambda=2\int^{+\infty}_{0}d\omega\frac{\alpha^{2}F\left(\omega\right)}{\omega}$. The correction functions are expressed by the formulas:

\begin{equation}
f_{1}=\left[1+\left(\frac{\lambda}{\Lambda_{1}}\right)^{\frac{3}{2}}\right]^{\frac{1}{3}}, \qquad f_{2}=1+\frac{\left(\frac{\sqrt{\omega_{2}}}{\omega_{\rm{ln}}}-1\right)\lambda^{2}}{\lambda^{2}+\Lambda^{2}_{2}}\nonumber,
\end{equation}
whereas: $\Lambda_{1}=2.46\left(1+3.8\mu^{\star}\right)$ and $\Lambda_{2}=1.82\left(1+6.3\mu^{\star}\right)\frac{\sqrt{\omega_{2}}}{\omega_{\ln}}$. 
The second moment of the normalized weight function should be calculated from:
$\omega_{2}=\frac{2}{\lambda}\int_{0}^{+\infty}d\omega\alpha^{2}F\left(\omega\right)\omega$.
The values of the remaining thermodynamic parameters can be calculated on the basis of the formulas presented in the papers \cite{Mitrovic1984A, Marsiglio1986A, Carbotte1990A}:
\begin{eqnarray}
\label{r4}
R_{\Delta}&=&\frac{2\Delta\left(0\right)}{k_{B}T_{c}}\\ \nonumber
&=&3.53\left[1+12.5\left(\frac{k_{B}T_{c}}{\omega_{\ln}}\right)^2\ln\left(\frac{\omega_{\ln}}{2k_{B}T_{c}}\right)\right],
\end{eqnarray}
\begin{eqnarray}
\label{r5}
R_{C}&=&\frac{\Delta C\left(T_{c}\right)}{C^{N}\left(T_{c}\right)}\\ \nonumber
&=&1.43\left[1+53\left(\frac{k_{B}T_{c}}{\omega_{\ln}}\right)^2\ln\left(\frac{\omega_{\ln}}{3k_{B}T_{c}}\right)\right],
\end{eqnarray}
and
\begin{eqnarray}
\label{r6}
R_{H}&=&\frac{T_{c}C^{N}\left(T_{c}\right)}{H^{2}_{c}\left(0\right)}\\ \nonumber
&=&0.168\left[1-12.2\left(\frac{k_{B}T_{c}}{\omega_{\ln}}\right)^2\ln\left(\frac{\omega_{\ln}}{3k_{B}T_{c}}\right)\right],
\end{eqnarray}
where: $\Delta C\left(T_{c}\right)=C^{S}\left(T_{c}\right)-C^{N}\left(T_{c}\right)$ is the specific heat jump at the critical temperature. In particular, the specific heat of the normal state should be calculated on the basis of the formula: 
$C^{N}\left(T\right)=\rho\left(0\right)\gamma k^{2}_{B}T$, where $\rho\left(0\right)$ is the value of the electron density of states on the Fermi level. The Sommerfeld constant is given by: $\gamma=\frac{2}{3}\pi^{2}\left(1+\lambda\right)$. Let us note that the formulas Eq.~\eq{r4}-Eq.~\eq{r6} will be denoted also as MZC, MC and C formula.  

\vspace*{0.5cm}

In the presented paper we have determined the thermodynamic properties of the phonon-induced superconducting state in vanadium. This state is characterized by the anomalously high value of $\mu^{\star}$. The detailed calculations were performed with the help of the Eliashberg equations and the semi-analytical formulas, which allowed us to compare both approaches. We have used the Eliashberg function derived from the work \cite{Wierzbowska2005A}, which was determined with the help of the DFT method. 

Let us note that the studies on the superconducting state in vanadium last since 1934. For the first time the sudden drop of the resistance to zero at temperature of $4.3$~K was discovered by Meissner and Westerhoff \cite{Meissner1934A}. 
In 1952, Wexler and Corak obtained $T_{c}=5.13$~K \cite{Wexler1952A}. Fourteen years later, Radebaugh and Keesom measured   $T_{c}=5.4$~K \cite{Radebaugh1966A}. Recent experiments showed comparable values of the critical temperature $T_{c}=5.3$~K \cite{Ishizuka1999A, Ishizuka2000A}. 

In 1982, Zasadzinski {\it et al.} presented the results for the superconducting state received as the part of the tunnelling experiment \cite{Zasadzinski1982A}. They found that the value of the electron-phonon coupling constant equals $0.83$, and they established the existence of the relatively weak depairing effects ($\mu^{\star}=0.15$). Hence, on the basis of the Allen-Dynes formula, the estimation of $T_{c}=6.2$~K \cite{Zasadzinski1982A} could be obtained. This value was slightly higher than that given by Radebaugher and Keesom, or the value measured in the tunnelling experiment by Westerdale ($T_{c}=5.36$~K) \cite{Westerdale2010A}. It should be emphasized that in 1996 first work \cite{Savrasov2008A} was published, which suggested the existence of very strong depairing effects in vanadium ($\mu^{\star}=0.3$), due to the calculated high value of the electron-phonon coupling constant ($\lambda=1.19$).

Very interesting results were obtained examining the superconducting state in vanadium, which was subjected to the influence of the high pressure. The experiment conducted in 2000 by Ishizuka {\it et al.} \cite{Ishizuka2000A}, in the range of the pressures up to $120$~GPa has shown the strong linear growth of the critical temperature's values together with the increasing applied pressure 
($\left[T_{c}\right]_{\rm max}=16.5$~K for $p=120$~GPa). Similar value of  $T_{c}$ was obtained by Louiss and Iyakutti in 2003 ($T_{c}=17.2$~K) \cite{Louis2003A}. So far the highest critical temperature equal to 
$T_{c}=25$~K was measured by Vaitheeswaran {\it et al.} for the pressure at $130$~GPa \cite{Vaitheeswaran2000A}.

\vspace*{0.5cm}

%
\begin{figure}
\includegraphics[width=\columnwidth]{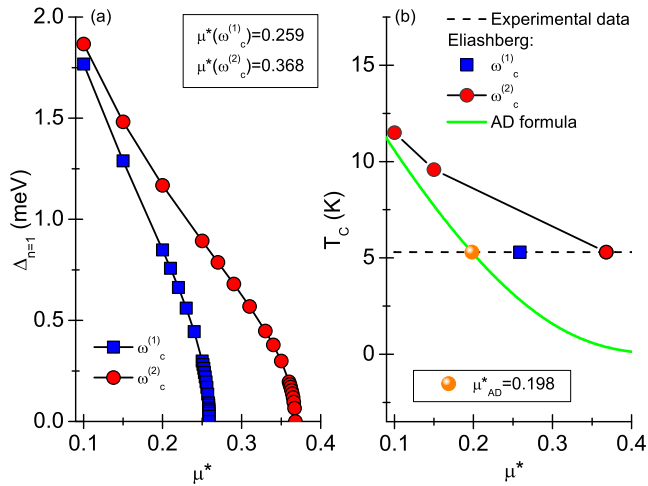}
\caption{(a) The dependence of the maximum value of the order parameter on the Coulomb pseudopotential for extreme $\omega_{c}$.  
             We have assumed $T_{c}=\left[T_{c}\right]_{\rm exp}$. 
         (b) The critical temperature as a function of $\mu^{\star}$. The results have been obtained with the use of the Eliashberg equations 
             and the Allen-Dynes (AD) formula.} 
\label{f1}
\end{figure}
%

In the first step we have determined the values of the Coulomb pseudopotential, for which the critical temperature calculated with the help of the Eliashberg equations reproduces the experimental results ($[T_{c}]_{\rm exp}=5.3$~K) \cite{Radebaugh1966A}. It turns out that the parameter 
$\mu^{\star}$ very strongly depends on the cut-off frequency. In the framework of the Eliashberg formalism it is generally accepted that 
$\omega_{c}\in\left<3\Omega_{\rm max}, 10\Omega_{\rm max}\right>$. Therefore, we have estimated the Coulomb pseudopotential for extreme points ($\omega_{c}^{\left(1\right)}=3\Omega_{\rm max}$ and $\omega_{c}^{\left(2\right)}=10\Omega_{\rm max}$). We have used the equation 
$\left[\Delta_{n=1}\left(\mu^{\star}\right)\right]_{T=T_{c}}=0$ due to the fact that $\Delta_{n=1}$ represents the maximum value of the order parameter on the imaginary axis. Of course, the Eliashberg equations set was solved in the self-consistent way, whereas we adopted $M=1100$. The numerical solutions of the Eliashberg equations were obtained with the use of the methods tested in the papers \cite{Ania2013A, Ania2014A, Ania2014B}.

The full dependence of the order parameter $\Delta_{n=1}$ on the Coulomb pseudopotential has been plotted in \fig{f1}~(a). It can be seen that, independently of adopted $\omega_{c}$ the value of the Coulomb pseudopotential is anomalously high 
($\mu^{\star}(\omega_{c}^{\left(1\right)})=0.259$ or $\mu^{\star}(\omega_{c}^{\left(2\right)})=0.368$). Physically it means that the depairing electron correlations in vanadium cannot be only linked with the Coulomb static screened repulsion corrected by the retardation effects. For example, in the work \cite{Wierzbowska2005A} the detailed attention is paid to the strongly destructive influence of the spin fluctuations. 

In \fig{f1}~(b) we have presented the dependence of the critical temperature on the Coulomb pseudopotential obtained with the help of the Eliashberg equations and the Allen-Dynes formula. It can be easily noticed that the semi-analytical approach gives the value of the Coulomb pseudopotential, which is much lower than the values obtained from the Eliashberg equations ($\mu^{\star}_{\rm AD}=0.198$). This is due to the approximations used in the course of deriving the Allen-Dynes formula, inter alia, the omission of the cut-off frequency in the analytical approach. Methodologically obtained result proves that quantity $\mu^{\star}$ in the Allen-Dynes formula should be regarded only as the fitting parameter, matching model to the experimental results. In fact, similarly we should look at the Coulomb pseudopotential in the Eliashberg equations in the case, when the value of $\mu^{\star}$ is far higher than $0.1$. This fact is due to the high volatility of the Coulomb pseudopotential's value together with the change of $\omega_{c}$. The physical meaning of the Coulomb pseudopotential's value can be expected only in the neighbourhood of the point $\mu^{\star}=0.1$, where both methods give comparable results (see \fig{f1}~(b) - results for $\mu^{\star}=0.1$ and $\mu^{\star}=0.15$). 

%
\begin{figure}
\includegraphics[width=\columnwidth]{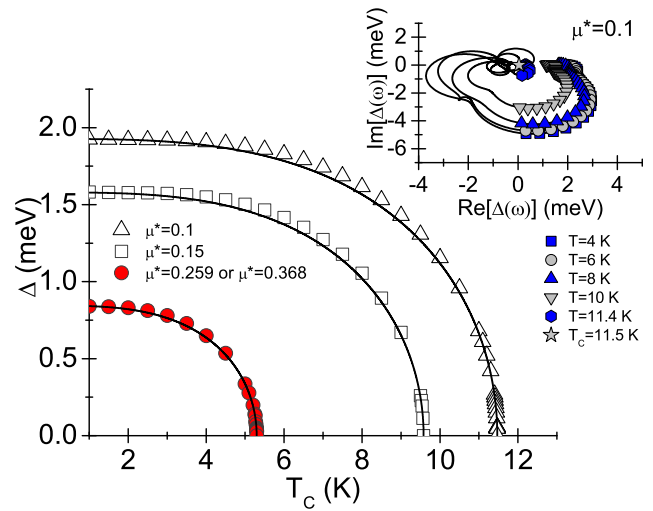}
\caption{The order parameter as a function of the temperature for the selected values of the Coulomb pseudopotential. The symbols have been obtained with the help of the Eliashberg equations, lines represent the values predicted by the BCS model: 
         $\Delta\left(T\right)=\Delta\left(0\right)\sqrt{1-\left(T\slash T_{c}\right)^{\Gamma}}$, where $\Gamma=3$ \cite{Eschrig2001A}.  
         The insert presents the exemplary courses of the order parameter on the complex plane in the dependence on the temperature.} 
\label{f2}
\end{figure}
%

In \fig{f2} we have posted the courses of the order parameter as a function of the temperature. Physical values of the order parameter were calculated on the basis of the equation $\Delta\left(T\right)={\rm Re}\left[\Delta\left(\omega=\Delta\left(T\right)\right)\right]$, while the order parameter on the real axis ($\Delta\left(\omega\right)$) was determined basing on the analytical continued method \cite{Beach2000A}. The exemplary values of the order parameter on the complex plane for the selected temperatures have been shown in the insertion in \fig{f2}. As expected, for the Coulomb pseudopotential equal to $\mu^{\star}(\omega_{c}^{\left(1\right)})$ or  $\mu^{\star}(\omega_{c}^{\left(2\right)})$ we have obtained the same function $\Delta\left(T\right)$. It differs only slightly from the shape of the curve of the BCS theory, which is the interesting result, since the electron-phonon coupling constant for vanadium is relatively high ($\lambda=0.91$). It appears that the observed result is due to the very high values of the Coulomb pseudopotential, which significantly weaken the strong-coupling and the retardation effects included in the Eliashberg formalism. The easiest way those effects can be modelled is to use the ratio $r=k_{B}T_{C}\slash \omega_{\rm ln}$. For $\mu^{\star}(\omega_{c}^{\left(1\right)})$ and $\mu^{\star}(\omega_{c}^{\left(2\right)})$ obtained $r=0.003$. In the case of the significantly lower values of the Coulomb pseudopotential we get 
$\left[r\right]_{\mu^{\star}=0.1}=0.006$ and  $\left[r\right]_{\mu^{\star}=0.15}=0.005$. Keep in mind that in the limit of BCS $r=0$ \cite{Carbotte1990A}. 

%
\begin{figure}
\includegraphics[width=\columnwidth]{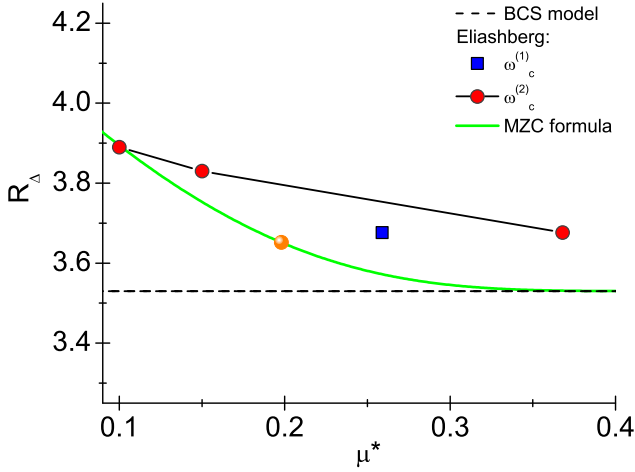}
\caption{The dimensionless ratio $R_{\Delta}$ as a function of the Coulomb pseudopotential. The results have been obtained on the basis 
         of the Eliashberg theory and the Mitrovic, Zarate, and Carbotte (MZC) formula.}
\label{f3}
\end{figure}
%

Having at disposal the results presented in \fig{f2} it is possible to compare the outcomes of the Eliashberg theory with the semi-analytical approach. For this purpose, we took into account the formula of Mitrovic, Zarate, and Carbotte, which allows to calculate the value of the dimensionless ratio 
$R_{\Delta}=2\Delta\left(0\right)\slash k_{B}T_{c}$ (see Eq.~\eq{r4}). The results are collected in \fig{f3}. 
Of course, for $\mu^{\star}(\omega_{c}^{\left(1\right)})$ and $\mu^{\star}(\omega_{c}^{\left(2\right)})$ in the framework of the full Eliashberg formalism we received the same result ($R_{\Delta}=3.67$), which is only slightly different from the result of the BCS theory ($R_{\Delta}=3.53$) \cite{Bardeen1957A, Bardeen1957B}. As it was mentioned, the lowering of the Coulomb pseudopotential's value strengthens the importance of the strong-coupling and the retardation effects, whereby the value of the parameter $R_{\Delta}$ increases ($\left[R_{\Delta}\right]_{\mu^{\star}=0.1}=3.89$ and $\left[R_{\Delta}\right]_{\mu^{\star}=0.15}=3.83$). The special attention should be paid to the fact that the MZC formula reproduces well the results of the advanced Eliashberg formalism, as long as we adopt $\mu^{\star}=\mu^{\star}_{\rm AD}$. In particular $\left[R_{\Delta}\right]_{\mu^{\star}_{\rm AD}}=3.65$.

%
\begin{figure}
\includegraphics[width=\columnwidth]{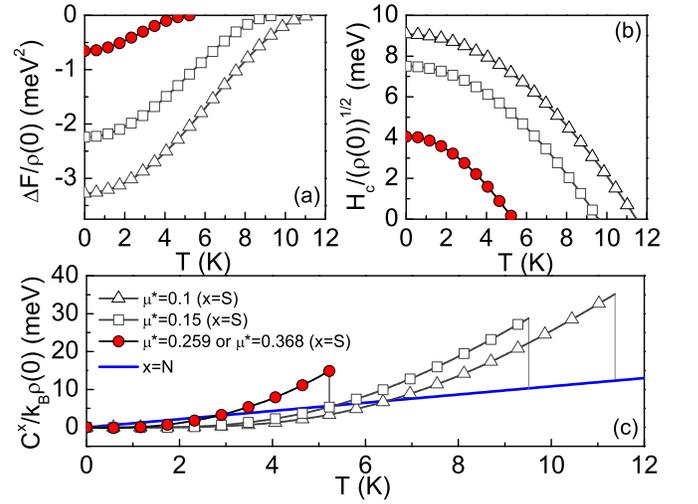}
\caption{(a) The free energy difference between the superconducting and the normal state. 
         (b) The thermodynamic critical field and (c) the specific heat of the superconducting state and the normal state.} 
\label{f4}
\end{figure}
%

%
\begin{figure}
\includegraphics[width=\columnwidth]{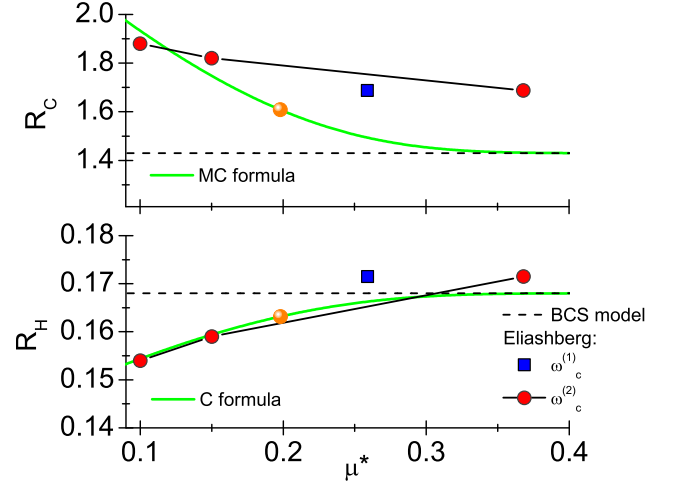}
\caption{(a) The values of $R_{C}$ parameter as a function of the Coulomb pseudopotential, (b) the values of $R_{H}$ parameter. 
         The results have been obtained with the help of the Eliashberg equations, the Marsiglio and Carbotte (MC) formula and the Carbotte (C) formula.} 
\label{f5}
\end{figure}
%

In the following part of the paper we have calculated the free energy difference between the superconducting and the normal state ($\Delta F$) \cite{Bardeen1964A}, the thermodynamic critical field, and the specific heat for the superconducting and the normal state. We have used the following formula:

\begin{eqnarray}
\label{r7}
\frac{\Delta F}{\rho\left(0\right)}&=&-\frac{2\pi}{\beta}\sum_{m=1}^{M}
\left(\sqrt{\omega^{2}_{m}+\Delta^{2}_{m}}- \left|\omega_{m}\right|\right)\\ \nonumber
&\times&(Z^{S}_{m}-Z^{N}_{m}\frac{\left|\omega_{m}\right|}
{\sqrt{\omega^{2}_{m}+\Delta^{2}_{m}}}),
\end{eqnarray}  
where $Z^{S}_{m}$ and $Z^{N}_{m}$ denote the wave function renormalization factor for the superconducting and the normal state. The thermodynamic critical field is given by the expression below:
\begin{eqnarray}
\label{r8}
\frac{H_{c}}{\sqrt{{\rho}(0)}}=\sqrt{-8\pi\left[\Delta F/\rho\left(0\right)\right]}.
\end{eqnarray}

The difference of the specific heat between the superconducting and the normal state can be determined with the help of the formula:
\begin{eqnarray}
\label{r9}
\frac{\Delta C\left(T\right)}{k_{B}\rho(0)}=-\frac{1}{\beta}\frac{d^{2}\left[\Delta F/\rho\left(0\right)\right]}{d\left(k_{B}T\right)^{2}}.
\end{eqnarray}

Obtained results have been collected in \fig{f4}. They allow to calculate the values of the dimensionless thermodynamic ratios 
$R_{C}=\Delta C\left(T_{c}\right)\slash C^{N}\left(T_{c}\right)$ and 
$R_{H}=T_{c}C^{N}\left(T_{c}\right)\slash H^{2}_{c}\left(0\right)$. Note that in the framework of the BCS theory the parameters $R_{C}$ and $R_{H}$ accept the universal values of $1.43$ and $0.168$, respectively \cite{Bardeen1957A, Bardeen1957B}. For vanadium, in case of 
$\mu^{\star}(\omega_{c}^{\left(1\right)})$ and $\mu^{\star}(\omega_{c}^{\left(2\right)})$ we have obtained $R_{C}=1.68$ and $R_{H}=0.171$. Of course, with increasing significance of the strong-coupling and the retardation effects due to decrease in the value of the Coulomb pseudopotential ($\mu^{\star}=0.1$ or $\mu^{\star}=0.15$) the difference between the predictions of the Eliashberg formalism and the BCS theory rises, which has also been presented in \fig{f5}. Additionally, it can be seen that the semi-analytical formula (Eq.~\eq{r5} and Eq.~\eq{r6}) reproduce relatively precisely the numerical results in the case, when $\mu^{\star}=\mu^{\star}_{AD}$. In particular, $\left[R_{C}\right]_{\mu^{\star}_{\rm AD}}=1.61$ and $\left[R_{H}\right]_{\mu^{\star}_{\rm AD}}=0.163$.  It should be emphasized that both approaches (Eliashberg's and semi-analytical) give the comparable results for the same values of the Coulomb pseudopotential near $\mu^{\star}=0.1$.

\vspace*{0.5cm}

Summarizing, we have determined the thermodynamic parameters of the superconducting state in vanadium. We have taken into account the approach based on the Eliashberg equations and the semi-analytical formulas. In spite of the fact that the electron-phonon coupling constant in vanadium assumes the relatively high value, determined thermodynamic functions differ little from their counterparts in the BCS theory. The obtained result is related to the fact that the value of the Coulomb pseudopotential (regardless of the cut-off frequency) is abnormally high, resulting in the significant reduction in the importance of the strong-coupling and the retardation effects. From the physical point of view, such the high value of the Coulomb pseudopotential means that it cannot be associated only with the Coulomb static screened repulsion corrected by the retardation effects. 

In addition, we have shown that the semi-analytical approach allows, with the good approximation, to reproduce the results of the advanced Eliashberg formalism in the case, when for $\mu^{\star}$ we accept the value from the Allen-Dynes formula. It should be emphasized that this value is significantly lower than the value of the Coulomb pseudopotential determined with the help of the Eliashberg equations, and it should be treated as the fitting parameter.

It also draws attention that both approaches give comparable results for the same value of the Coulomb pseudopotential only near point $\mu^{\star}=0.1$. In the case under consideration, the parameter $\mu^{\star}$ is the measure of the Coulomb static screened repulsion corrected by the retardation effects. Generally speaking, however, going beyond the description of the depairing electron correlations with the help of the Coulomb pseudopotential seems to be the very interesting direction of research. Particularly, in the context of the superconducting systems, in which induce the electron phases that are competitive with the superconducting state \cite{Balseiro1979A, Szczesniak2016A} or we expect the strong electron correlations \cite{Bednorz1986A, Bednorz1988A}. Work in this direction has been taken by us.

\bibliography{Bibliography}
\end{document}